\documentclass[nofootinbib,superscriptaddress,a4paper,twocolumn, floatfix]{revtex4-2}
\usepackage[english]{babel}
\usepackage[utf8]{inputenc}

\usepackage{amsthm}
\usepackage{mathtools}
\usepackage{physics}
\usepackage{xcolor}
\usepackage{graphicx}
\usepackage[left=23mm,right=13mm,top=35mm,columnsep=15pt]{geometry} 

\usepackage{algorithm2e}
\usepackage{algpseudocode}
\makeatletter
\newenvironment{Method}[1][htb]{%
   \begin{procedure}[#1]%
  }{\end{procedure}}
\makeatother

\usepackage{adjustbox}
\usepackage{placeins}
\usepackage[T1]{fontenc}
\usepackage{lipsum}
\usepackage{csquotes}
\usepackage[pdftex, pdftitle={Article}, pdfauthor={Author}]{hyperref}
\usepackage{xcolor}
\hypersetup{
    colorlinks,
    linkcolor={red!50!black},
    citecolor={blue!50!black},
    urlcolor={blue!80!black}
}
\usepackage{booktabs}
\usepackage{multirow}
\usepackage{listings}
\usepackage{xcolor}

\definecolor{codegreen}{rgb}{0,0.6,0}
\definecolor{codegray}{rgb}{0.5,0.5,0.5}
\definecolor{codepurple}{rgb}{0.58,0,0.82}
\definecolor{tqblue}{HTML}{08293d}
\definecolor{backcolour}{HTML}{fefdf5}

\lstdefinestyle{mystyle}{
    backgroundcolor=\color{backcolour},   
    commentstyle=\color{codegreen},
    keywordstyle=\color{magenta},
    numberstyle=\tiny\color{codegray},
    stringstyle=\color{codepurple},
    basicstyle=\ttfamily\footnotesize\color{tqblue},
    breakatwhitespace=false,         
    breaklines=true,
    postbreak=\mbox{\textcolor{magenta}{$\hookrightarrow$}\space},                 
    captionpos=b,                    
    keepspaces=true,                 
    numbers=left,                    
    numbersep=5pt,                  
    showspaces=false,                
    showstringspaces=false,
    showtabs=false,                  
    tabsize=2
}

\newcommand{\shiftleft}[2]{\makebox[0pt][r]{\makebox[#1][l]{#2}}}

\begin{document}

\title{A Quantum Algorithmic Approach to Multiconfigurational Valence Bond Theory: Insights from Interpretable Circuit Design}

\author{Jakob~S.~Kottmann}
\email[E-mail:]{jakob.kottmann@uni-a.de}
\affiliation{{Institute for Computer Science, University of Augsburg, Germany }}

\author{Francesco~Scala}
\affiliation{{Dipartimento di Fisica, Universit\`{a} degli Studi di Pavia, Italy}} 

\date{\today} 
\begin{abstract}
Efficient ways to prepare fermionic ground states on quantum computers are in high demand and different techniques have been developed over the last years. Despite having a vast set of methods, it is still unclear which method performs well for which system.
 In this work, we combine interpretable circuit designs with an effective basis approach in order to optimize a multiconfigurational valence bond wavefunction. Based on selected model systems, we show how this leads to explainable performance.
    We demonstrate that the developed methodology outperforms related methods in terms of the size of the effective basis as well as individual quantum resources for the involved circuits. 
\end{abstract}

\maketitle

\phantom{a}\\
Over the last few years, multiple ground-state methods for many-body systems
were designed as hybrid approaches for quantum and classical computers leveraging effective bases. In such scenarios,
a matrix representation of the original Hamiltonian H
is computed in a basis of
qubit states $\ket{\psi_k}$ generated by the unitaries $U_k$. As this basis is usually not orthogonal, the generalized eigenvalue
equation
\begin{align}
    \mathbf{H} \mathbf{c} = \lambda \mathbf{S} \mathbf{c}, 
    \;\;H_{ij} = \bra{\psi_i}H\ket{\psi_j},\;\; S_{ij} = \braket{\psi_i}{\psi_j},\label{eq:effective-diagonalization}
\end{align}
is solved with Hamiltonian and overlap matrix elements measured on the qubit device.
This strategy is often introduced as an alternative to variational methods~\cite{bharti2022noisy, cerezo2021variational}, in particular the variational quantum eigensolver~\cite{mcclean2016theory, peruzzo2014variational}, currently representing the largest class of algorithmic procedures on hybrid hardware.
Techniques, such as in Refs.~\citenum{anand2022quantum, tilly2020computing, lee2018generalized, kottmann2020reducing, kottmann2021feasible}, developed in the context of variational methods can however be employed within the effective diagonalization of Eq.~\eqref{eq:effective-diagonalization} as well.\\

\begin{figure*}
\includegraphics[width=0.95\textwidth]{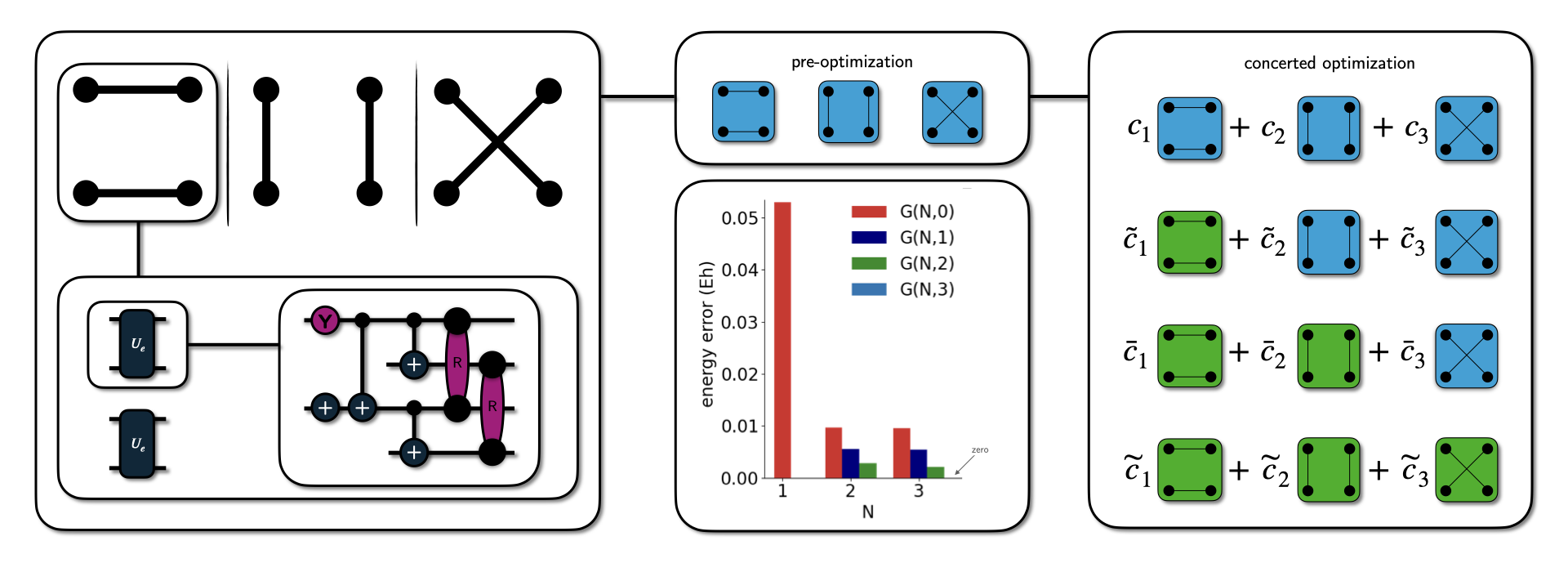}
\caption{Basis construction: Illustration for the construction of many-body basis states through graph-based heuristics following Ref.~\citenum{kottmann2022molecular}. On the left: the graphs with one of the corresponding SPA~\cite{kottmann2022optimized, kottmann2022molecular} circuits of the {square} H$_4$/STO-6G(4,8) constructed via Heuristic~1 in Ref.~\citenum{kottmann2022molecular}. On the lower center: Energetic errors with different levels of optimization. $G(N,M$$\leq$$N)$ (see Method 1) the optimization~\eqref{eq:concerted-opt} of parameters from $M$ circuits in a total  wavefunction assembled from $N$ circuits as in Eq.~\eqref{eq:wavefunction}.
On the right: illustration of the individual $G(N,M)$ wavefunctions obtained through concerted optimizations of the coefficients and angles from $M$ circuits with the corresponding graphs highlighted in green, {angles form graphs highlighted in blue are fixed at values from the pre-optimization}. From top to bottom the illustrations correspond to $G(3,0), G(3,1), G(3,2), G(3,3)$.
}
\label{fig:graph_basis_construction}
\end{figure*}

The generation of the effective basis can be broadly divided into two classes of methods, both starting with a suitable initial  state $\ket{\psi_0}$ -- a state with at least non-vanishing, but ideally high, overlap with the ground state of the Hamiltonian of interest, which is then used to construct the basis over a set of unitary operations $\ket{\psi_k} = U_k\ket{\psi_0}$.
The first class comes in the form of a \textit{non-orthogonal variational quantum eigensolver} (NOVQE)~\cite{huggins2020non} with pre-trained many-body basis states constructed from variational quantum circuits 
\begin{align}
    &\ket{\psi_k} = U_k(\boldsymbol{\theta}_k^{(*)})\ket{\psi_0}\\
    &\boldsymbol{\theta}^{(*)}_k = \text{argmin}_{\boldsymbol{\theta}_k} \expval{H}_{U_k\left(\boldsymbol{\theta}_k\right)}\vphantom{\sum_a^b},\label{eq:novqe}
\end{align}
{typically with $U_k$ being of unitary coupled-cluster type and $\ket{\psi_0}$ the Hartree-Fock state.}
The other class~\cite{stair2020multireference,yeter2020practical,motta2020determining,tsuchimochi2022multistate,parrish2019quantum2,cohn2022quantum,bharti2021quantum, bharti2021quantum, bharti2021iterative,kyriienko2020quantum,seki2021quantum,kirby2023exact}, uses non-variational unitaries, usually derived from the Krylov subspace
\begin{align}
    \mathcal{K} = \left\{ U_k\ket{\psi_0} \propto H^k\ket{\psi_0} \right\}_{k=1}^{N},\label{eq:krylov}
\end{align}
where the details lie in the methodologies to approximate $\mathcal{K}$ (see appendix~\ref{sec:related-works} for more). {While it is, in principle, clear how to construct the Krylov basis, the hope of NOVQEs and related methods is to find better effective bases to describe the problem at hand. Here, most methods so far resort to generic circuit designs from unitary coupled-cluster~\cite{anand2022quantum} preventing a deeper insight, that more interpretable methods can potentially offer. }\\
 
Recent examples of such endeavors are for example graph based representations of quantum states, either through interpretable quantum circuit design~\cite{kottmann2022molecular} or in the context of quantum optical setups~\cite{krenn2021conceptual,kottmann2021quantum,ruiz2022digital,arlt2022digital}, and concepts in quantum machine learning~\cite{schuld2021effect,anand2022information, heese2023explainable, casas2023multidimensionala}. Such techniques are not only useful for more effective computational protocols, but their true strength lies in their interpretability allowing for the extraction of principles and insights from small numerical computations that can be leveraged to tackle larger computational tasks more effectively.  \\

In this work, we use the interpretable circuit design of Ref.~\citenum{kottmann2022molecular} in order to determine compact effective bases suitable to capture the essential physics of a fermionic ground state. We show how this can be leveraged to gain insight from numerical results leading to explainable concepts of the missing effects for a full description of the ground state of interest. This, for example, gives an intuitive explanation of why energy based pre-optimization in the style of Eq.~\ref{eq:novqe} often fails which is illustrated with a detailed example.\\

{The construction of the wavefunctions in this work is closely related to multiconfigurational valence bond theory with optimized orbitals (often referred to as VBSCF)~\cite{hiberty2007textbook9} which is exact for the chosen space of active orbitals if all valence bond structures are included. The computational bottleneck of this method is the evaluation of the effective Hamiltonian matrix -- in this work, this evaluation is performed on the quantum processor with orbital rotations explicitly represented by unitary circuits. Detailed comparison with respect to potential computational advantages are out-of-scope for this work, we however think that this an interesting link between Valence Bond Theory and Quantum Computing approaches worth pointing out. In this work we develop  the basic technology for the implementation accompanied with initial applications and examples aimed at providing a deeper insight into the methodology. In order to ensure a solid foundations for future improvements we provide an open-source implementation within the \textsc{tequila}~\cite{tequila} package. Note that the methodology developed in this work is different to other approaches related to valence bond methods, such as in Ref.~\citenum{ghasempouri2023modular} (with connections to Resonating Valence Bond Theory) or Ref.~\citenum{kottmann2022optimized} (with connections to Generalized Valence Bond Theory).}\\

We will begin by providing more details on the method in Sec.~\ref{sec:method} followed by an overview over the used circuit construction methods~\ref{sec:circuits}. 
In Sec.~\ref{sec:results} we illustrate the performance of the developed techniques on explicit use cases that were used in previous works. Here we provide a detailed analysis using a prominent benchmark system and illustrate with extended numerical simulations that our method results in compacter bases with respect to basis size and cost of the individual circuits.

\section{Method}\label{sec:method}

We start by selecting a suitable many-body basis in the form of parametrized quantum circuits
\begin{equation}
    \label{eq:single wf}
\ket{\psi\left(\boldsymbol{\theta}_k\right)} = U_k\left(\boldsymbol{\theta}_k\right)\ket{0},
\end{equation}
in order to represent the total wavefunction
\begin{align}
    \ket{\Psi\left(\mathbf{c},\boldsymbol{\theta}\right)} = \sum_{k=1}^{N} c_k U_k \left(\boldsymbol{\theta}_k\right)\ket{0}.\label{eq:wavefunction}
\end{align}
with $\ket{0} \equiv \ket{00\dots0}$ denoting the fermionic vacuum state (all qubits in zero). 
Depending on the ground state problem of interest, the choice of the circuits $U_k$ will have practical implications on the runtime of the involved optimizations and the quality of the final wavefunction. In Sec.~\ref{sec:circuits} we will discuss a specific choice of circuits suitable for electronic structure, which we apply in this work. {Note that the wavefunction in Eq.~\eqref{eq:wavefunction}, constructed by  circuits $U_k$ described in the next section, fulfills all requirement of a Valence Bond wavefunction (\textit{cf.} Eqs.~(1) and~(9) in Ref.~\citenum{wu2011classicalvbt}).}\\

Once the basis elements are chosen, a concerted optimization of all parameters in the total wavefunction Eq.~\eqref{eq:wavefunction} is performed
\begin{align}
    \left\{\mathbf{c}^{(*)},\boldsymbol{\theta}^{(*)}\right\} = \underset{{\boldsymbol{c}, \boldsymbol{{
    \theta}}^{(M)}}}{\operatorname{argmin}}  \frac{\langle \Psi\left(\mathbf{c},\boldsymbol{\theta}\right) \rvert H \lvert \Psi\left(\mathbf{c},\boldsymbol{\theta}\right) \rangle} {\langle \Psi\left(\mathbf{c},\boldsymbol{\theta}\right) \vert \Psi\left(\mathbf{c},\boldsymbol{\theta}\right)\rangle},
    \label{eq:concerted-opt}
\end{align}
with $\boldsymbol{{\theta}}^{(M)} = \bigcup_{k=1}^{M}\left\{\boldsymbol{\theta}_k\right\}$ denoting the set of parameters subjected to the optimization procedure.
Depending on the wavefunction and parameters in the optimization we {describe the effective basis method using} the notation $G(N,M)$ with $N$ denoting the number of circuits included in Eq.~\eqref{eq:wavefunction} and $M\leq N$ the number of fully optimized parameters $\boldsymbol{{\theta}}^{(M)}$.
After the concerted optimization the generalized eigenvalue equation~\eqref{eq:effective-diagonalization} is invoked as a convergence test. If the so-determined coefficients $\boldsymbol{c}$ differ from the optimized coefficients ${\boldsymbol{c}^{(*)}}$, the concerted optimization is restarted with the coefficients determined through Eq.~\eqref{eq:effective-diagonalization} as starting values. This scheme proved to be useful in similar optimization methods.~\cite{kottmann2015accurate}\\

Prior to the optimization in Eq.~\eqref{eq:concerted-opt}, the circuit parameters can be initialized in the spirit of NOVQE~\cite{huggins2020non} through individual energy optimization as in Eq.~\eqref{eq:novqe}. {In addition, notice that $G(N,0)$ is equivalent to NOVQE and $G(1,1)=G(1,0)$ is equivalent to SPA.}

At this point, it is crucial to avoid linear dependencies (i.e. restricting the overlaps $S_{ij}$ from becoming too close to one).~\cite{epperly2022theory} This can either be done by including penalty terms into the optimization or through the design of the individual circuits -- in this work we resort to the latter (see Sec.~\ref{sec:circuits}) and provide an argument (Sec.~\ref{sec:results}) why this can be advantageous. The pre-optimized circuits are then subjected to the generalized eigenvalue equation~\eqref{eq:effective-diagonalization} resulting in initial values for the coefficients $\mathbf{c}^\text{(0)}$ and initial energies which we will denote as $G(N,0)$. 

{
In Method~1 the key aspects of this section are summarized and an explicit implementation can be found in the appendix~\ref{sec:code-example}.
}

\begin{Method}
\caption{G(N,M)}
\begin{algorithmic}
    \State 1. choose a suitable collection of circuits $U_k$
    \State 2. select $N$ circuits
    \State 3. pre-optimize  $\boldsymbol{\theta}_k^0 = \operatorname{argmin}_{\boldsymbol{\theta_k}} \expval{H}_{U_k\ket{0}}$
    \State 4. initialize $\boldsymbol{\theta} = \bigcup \left\{\boldsymbol{\theta}_k^0\right\}$
    \State 5. compute initial coefficients $\boldsymbol{c}^0$ via Eq.~\eqref{eq:effective-diagonalization}
    \State 6. select $M\leq N$ circuits
    \State 7. initialize $\boldsymbol{\theta}^{(M)} = \bigcup_{k \in M} \left\{\boldsymbol{\theta}_k^0\right\}$
    \State 8. optimize $\boldsymbol{c},\boldsymbol{\theta}^{(M)}$ parameters via Eq.~\eqref{eq:concerted-opt}
    \State 9. check convergence via Eq.~\eqref{eq:effective-diagonalization}
    \State 10. repeat until converged
\end{algorithmic}
\end{Method}

\section{Circuits}\label{sec:circuits}
In this work we will employ the circuit design principles of Ref.~\citenum{kottmann2022molecular} in the context of fermionic Hamiltonians (encoded via Jordan-Wigner) in the usual form
\begin{align}
    H_\text{f} = \sum_{kl} h^l_k a^\dagger_k a_l + \frac{1}{2}\sum_{klmn} g^{mn}_{kl} a^\dagger_k a^\dagger_l a_n a_m,\label{eq:hamiltonian}
\end{align}
with fermionic annihilation ($a_i$) and creation operators ($a^\dagger_i$) {that annihilate or create} electrons in the one-body basis state (also called spin-orbital) $\ket{\chi_i}$. {In the text, we will follow the convention that $2N_\text{O}$ spin-orbitals are constructed from $N_\text{O}$ spatial orbitals $\ket{\phi_{k}}$ as 
$
    \ket{\chi_{2k}} = \ket{\phi_k} \otimes \ket{\uparrow},
    \ket{\chi_{2k+1}} = \ket{\phi_k} \otimes \ket{\downarrow}.
$
Note however, that other orderings are possible (e.g. $\uparrow \uparrow \dots \uparrow \downarrow\downarrow\dots \downarrow$ used in the toc graphic).} In the molecular case, the tensors $h$ (one body integrals) and $g$ (electronic repulsion integrals) are computed as integrals over the states $\chi_i$ (for details see Ref.~\citenum{kottmann2020reducing} or the appendix A of Ref.~\citenum{kottmann2022molecular}).
In this section we will resort to a short illustrative summary of the applied circuit designs.\\

Assume that we have a fermionic Hamiltonian as in Eq.~\eqref{eq:hamiltonian} with $N_O$ spatial orbitals, and a collection of graphs, with vertices $V$ corresponding to sets of uniquely assigned spatial orbitals and a set of non-overlapping (no shared vertices) edges $E$. We can then construct a quantum circuit from {a graph $G$ with edges $E$} as
\begin{align}
U_\text{G} = \bigotimes_{e\in E} U_e(\boldsymbol{\theta}_e)\label{eq:graph_circuit}  
\end{align}
where the individual circuits $U_e(\boldsymbol{\theta}_e)$ prepare a two-electron wavefunction on the qubits corresponding to the edge. The total wavefunction is then a $2|E|$-electron wavefunction and corresponds to a separable pair approximation (SPA)~\cite{kottmann2022optimized} (see Heuristic~1 and the appendix~G in Ref.~\citenum{kottmann2022molecular} for more details {and a comment on odd-electron numbers}).
The circuits $U_e$ for the simplest non-trivial {case, namely,} one orbital assigned to each vertex (corresponding to two spin-orbitals or qubits) are illustrated in Fig.~\ref{fig:graph_basis_construction} and consist of two parametrized parts ($\boldsymbol{\theta}_e = \left\{\theta, \varphi\right\}$)
\begin{align}
U_e =  U_\text{R}\left(\varphi\right) U\left(\theta\right).\label{eq:edge_circuit}
\end{align}
The term $U(\theta)$ is built from a parametrized $R_y$ rotation and three controlled-NOT operations preparing the 4-qubit wavefunction 
\begin{align}
U\left(\theta\right)\ket{0} =& \cos\left(\frac{\theta}{2}\right)\ket{1100} + \sin\left(\frac{\theta}{2}\right)\ket{0011} \nonumber \\
\equiv&  \cos\left(\frac{\theta}{2}\right) \ket{\raisebox{-0.25cm}{\includegraphics[width=0.04\textwidth]{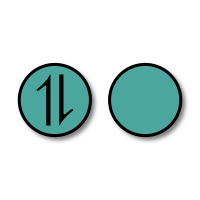}}} +  \sin\left(\frac{\theta}{2}\right) \ket{\raisebox{-0.25cm}{\includegraphics[width=0.04\textwidth]{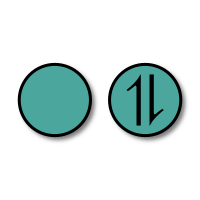}}},
\end{align}
where we used a graphical notation illustrating the two electrons in one of the two spatial basis states of the left and right vertex that form the edge $e$. As illustrated, $U(\theta)$ 
is moving a quasi-particle (often referred to as hard-core Boson)~\cite{kottmann2022optimized, elfving2021simulating} of two spin-paired electrons through the spatial basis states. However, the latter are not unique and we can transform the spatial part of the basis via linear combinations of the basis states.
The second part of the circuit $U_R(\varphi)$ implements a unitary operation that corresponds to such a basis change (see Ref.~\citenum{kottmann2022molecular} appendix B or Ref.~\cite{anselmetti2021local} for details). Graphically, this can be illustrated as
\begin{align}
U_\text{R}\left(\varphi\right)\ket{\raisebox{-0.25cm}{\includegraphics[width=0.04\textwidth]{configuration.001.png}}}
\equiv& \ket{\raisebox{-0.25cm}{\includegraphics[width=0.04\textwidth]{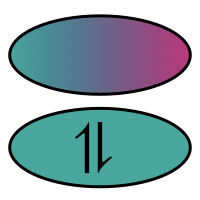}}} \label{eq:edge-bond} 
\end{align}
where the colors indicate positive and negative interference in the linear combination of the two basis states. In particular, the right hand side of Eq.~\eqref{eq:edge-bond} re-expresses the transformed wavefunction in the spatial orbitals 
\begin{align}
    \ket{\tilde{\phi}_1} &= \cos(\varphi/2) \ket{\phi_1} + \sin(\varphi) \ket{\phi_2},\\\ket{\tilde{\phi}_2} &= \cos(\varphi/2) \ket{\phi_1} - \sin(\varphi/2) \ket{\phi_2}.
\end{align}

Note that, in contrast to other works~\cite{kottmann2022optimized, elfving2021simulating, khamoshi2022agpbased,zhao2023orbital} we are not restricting the total Hamiltonian to the hard-core Boson approximation, but rather express the total wavefunction as a linear combination of hard-core Bosonic wavefunctions in different orbital bases.

\section{Results}\label{sec:results}
In the following we are applying the developed method to standard molecular benchmark systems used in Refs.~\citenum{stair2020multireference, huggins2020non, kottmann2022molecular} where we are interested mainly in the sizes of the effective bases sufficient to describe the electronic ground states. {We will denote the molecules as \textit{name/basis}($N_e$, \text{2}$N_\text{O}$) indicating the number of active electrons $N_e$ and active spin orbitals 2$N_{\text{O}}$ (usually corresponding to the number of qubits).} Through the circuit design described in the previous section, we are able to gain insights into the nature of the ground states. {For the hydrogen chains we fixed the H-H distance to 1.5 \AA, as this bond distance is challenging for an SPA circuit (see Fig.~\ref{fig:h6-spa}), by comparison with results from the literature (e.g. Fig.~4 in Ref.~\citenum{yordanov2021qubit} or Tab.~2 in Ref.~\citenum{kottmann2022molecular}) the SPA circuit provides a good baseline for the effective basis.} \\

The presented data is generated with \textsc{tequila}~\cite{tequila,kottmann2021feasible}, where we give explicit code examples in the appendix~\ref{sec:code-example}. In the computational process, the following dependencies were utilized in the background: \textsc{qulacs}~\cite{qulacs} as quantum backend, BFGS implementation within \textsc{scipy}~\cite{scipy} as optimizer, \textsc{pyscf}~\cite{pyscf1,pyscf2} to compute molecular integrals (forming the tensors of the Hamiltonian in Eq.~\eqref{eq:hamiltonian}) and exact energies, and the Jordan-Wigner encoding from~\textsc{openfermion}~\cite{OpenFermion}. The MRSQK energies were computed with \textsc{qforte}~\cite{stair2022qforte}.

\subsection{The H$_4$ Square System: insights from an explicit example}
In Fig.~\ref{fig:graph_basis_construction} the $G(N,M)$ method is illustrated on the H$_4$/STO-6G(4,8) square system consisting of four hydrogen atoms, each equipped with a single orbital from the STO-6G set, equidistantly placed on a rectangle with vertex distances set to $d=1.5$ \AA.
{Following Ref.~\citenum{kottmann2022molecular} we can construct three possible molecular graphs for this molecule, by assigning the 4 atomic basis orbitals to 4 edges and building the three perfect matchings of the fully connected graph. The corresponding graphs are illustrated in Fig.~\ref{fig:graph_basis_construction}.
The three graphs give rise to three circuits via Eqs.~\eqref{eq:graph_circuit} and ~\eqref{eq:edge_circuit}, that are used as effective basis in the wavefunction of Eq.~\eqref{eq:wavefunction}. The left panel of Fig.~\ref{fig:graph_basis_construction}  illustrates this circuit construction. Through the optimization procedure described in the previous section optimal circuit parameters and energies are determined.
In the central panel of Fig.~\ref{fig:graph_basis_construction} the resulting energies G($N$,$M$) with varying $N$ and $M$ are depicted.
}
A concerted optimization over all three graphs $G(3,3)$ yields the exact energy. There is however no visible difference between $G(2,M)$ and $G(3,M)$ with $M\neq3$ meaning that inclusion of the third graph only has an effect in a concerted optimization.

The wavefunction prepared by the circuit derived from the third graph state optimized on G(3,3) level -- \textit{i.e.} in a concerted optimization including all three graphs -- takes the form 
\begin{align}
\ket{\raisebox{-0.25cm}{\includegraphics[width=0.04\textwidth]{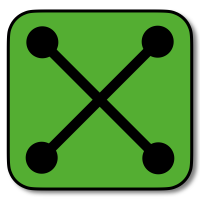}}}
 = \frac{1}{2}\left( 
  \ket{\raisebox{-0.25cm}{\includegraphics[width=0.04\textwidth]{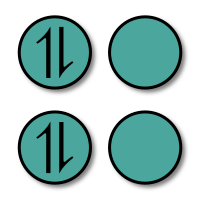}}} + 
  \ket{\raisebox{-0.25cm}{\includegraphics[width=0.04\textwidth]{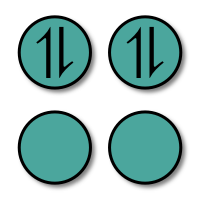}}} - 
  \ket{\raisebox{-0.25cm}{\includegraphics[width=0.04\textwidth]{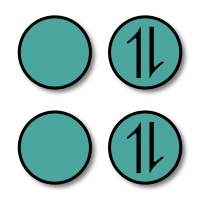}}} - 
  \ket{\raisebox{-0.25cm}{\includegraphics[width=0.04\textwidth]{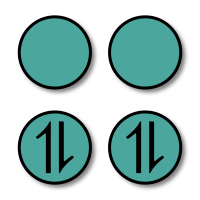}}}
 \right),\label{eq:graphstate.crossed}
\end{align}
Here the up (and down) arrows represent a spin-up (spin-down) electron occupying one of the four spatial orbitals located on the hydrogen atoms. The state in Eq.~\eqref{eq:graphstate.crossed} is assembled from four configurations that cluster the four electrons as close as possible. The wavefunction clearly is energetically not favorable explaining the failure of energy based pre-optimization in the $G(3,M<3)$ wavefunctions.
The reason why the state in Eq.~\eqref{eq:graphstate.crossed} needs to take this specific form becomes clear when we take a look at the $G(2,2)$ wavefunction

\begin{align}
\ket{G(2,2)} =&  \bar{c}_1 \ket{\raisebox{-0.25cm}{\includegraphics[width=0.04\textwidth]{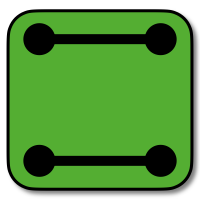}}} + \bar{c}_2 \ket{\raisebox{-0.25cm}{\includegraphics[width=0.04\textwidth]{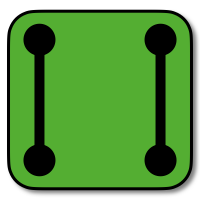}}}
 \label{eq:G(2,2)-wfn} \nonumber\\ 
  =& a\left(  
\ket{\raisebox{-0.25cm}{\includegraphics[width=0.04\textwidth]{configuration.008.png}}} + 
\ket{\raisebox{-0.25cm}{\includegraphics[width=0.04\textwidth]{configuration.009.png}}} - 
\ket{\raisebox{-0.25cm}{\includegraphics[width=0.04\textwidth]{configuration.010.png}}} - 
\ket{\raisebox{-0.25cm}{\includegraphics[width=0.04\textwidth]{configuration.011.png}}} \right)\nonumber\\
 & + b\ket{\Psi_\text{R}} \nonumber \\ 
 =& 2a \ket{\raisebox{-0.3cm}{\includegraphics[width=0.045\textwidth]{graph-crossed-green.png}}} +  b\ket{\Psi_\text{R}}
\end{align}
with the amplitudes $\bar{c}_2 = -\bar{c}_1$ and $a\approx 0.07 < b$ and $\Psi_\text{R}$ contains all other electronic configurations. We see from Eq.~\eqref{eq:G(2,2)-wfn} that the (optimized) wavefunction of the third graph Eq.~\eqref{eq:graphstate.crossed} is already included. Adding the third graph to the total wavefunction in $G(3,3)$ does therefore not introduce new configurations into the total wavefunction but it allows a relative reduction of the amplitude $a$ while preserving the internal structure of the residual (and energetically more important) wavefunction $\ket{\Psi_\text{R}}$ -- the structure of the $G(2,2)$ wavefunction alone would not allow this. On the other hand, energy based pre-optimization of the third graph does not result in the energetically unfavorable form of Eq.~\eqref{eq:graphstate.crossed} leading to $G(3,2)$ having no visible improvement over $G(2,2)$ as witnessed in Fig.~\ref{fig:graph_basis_construction}.
The analysis of the wavefunctions in Eqs.~\eqref{eq:graphstate.crossed} and~\eqref{eq:G(2,2)-wfn} also shows why orthogonality constraints between the individual graphs in the optimization can become problematic. \\

When compared to two isolated H$_2$ molecules the optimal wavefunction of the third graph would correspond to a product of two ionic non-bonding states, as it can be written as
\begin{align}
\ket{\raisebox{-0.25cm}{\includegraphics[width=0.04\textwidth]{graph-crossed-green.png}}} 
= \frac{1}{2}
\left(
\ket{\raisebox{-0.25cm}{\includegraphics[width=0.04\textwidth]{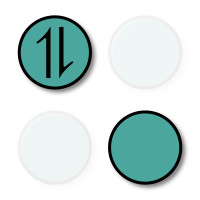}}}-
\ket{\raisebox{-0.25cm}{\includegraphics[width=0.04\textwidth]{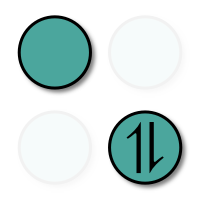}}}
\right) \otimes \left(
\ket{\raisebox{-0.25cm}{\includegraphics[width=0.04\textwidth]{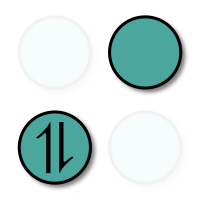}}}-
\ket{\raisebox{-0.25cm}{\includegraphics[width=0.04\textwidth]{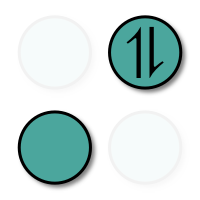}}}
\right)
\end{align}
while the corresponding wavefunctions of the other two graphs have more similarity with a product of bonding H$_2$ wavefunctions. The intuitive picture of the $G(2,2)$ wavefunction in Eq.~\eqref{eq:G(2,2)-wfn} that represents the H$_4$ wavefunction as a superposition of both degenerate realizations of two individual H$_2$ molecules is therefore still a reasonable model for the true ground state of the system. A suitable interpretation for the third graph is the addition of weak correlation between the two isolated H$_2$ molecules achieved by destructive interference of energetically unfavorable configurations. This allows an interesting connection to Ref.~\citenum{krenn2021conceptual} (in particular Fig.~4) where similar effects were identified in the context of quantum optical setups and similar arguments as in this case will hold for potential future approaches based on individual optimization.\\

\begin{figure*}
\includegraphics[width=0.32\textwidth]{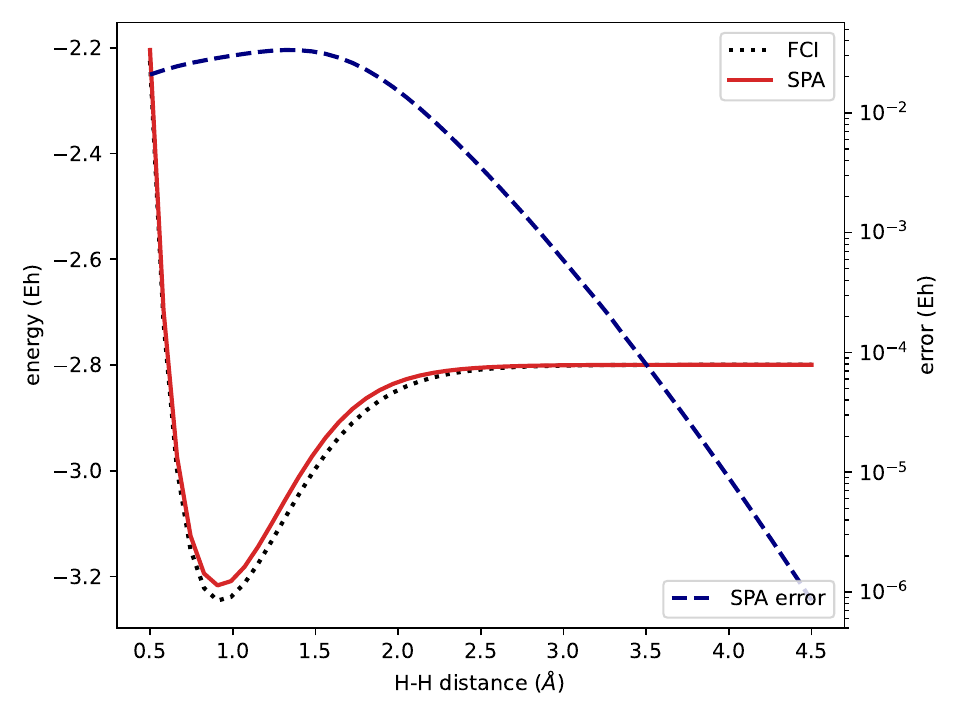}
\shiftleft{4.5cm}{\raisebox{2.5cm}[0cm][0cm]{
\includegraphics[width=0.09\textwidth]{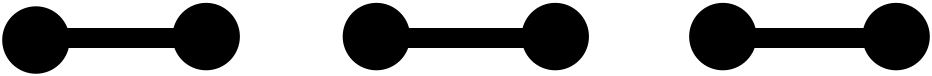}
}}
\includegraphics[width=0.32\textwidth]{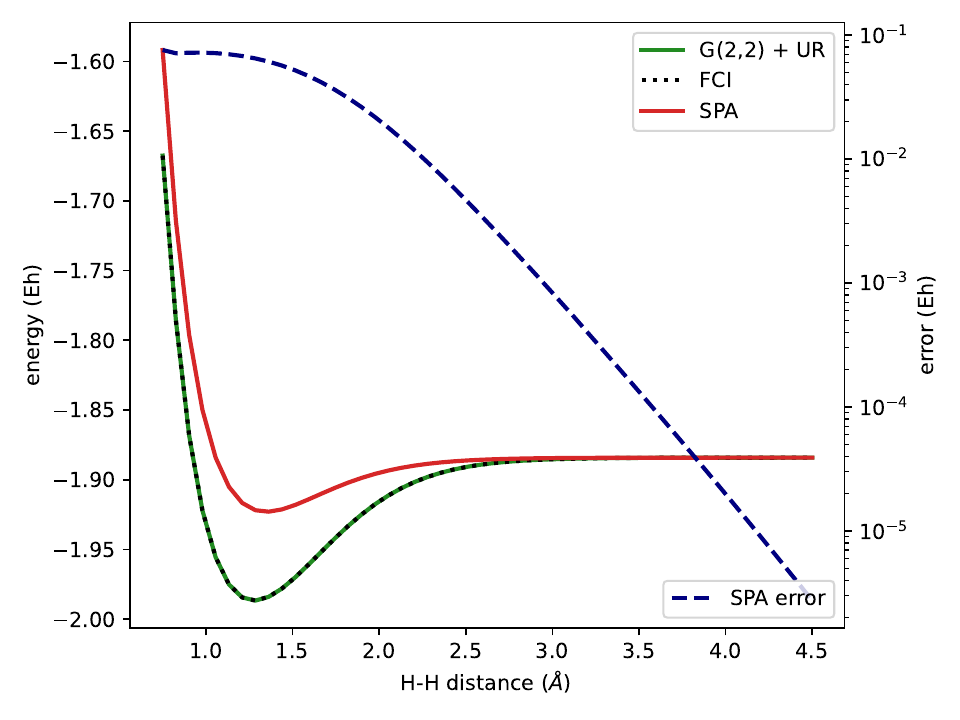}
\shiftleft{4.5cm}{\raisebox{1.8cm}[0cm][0cm]{
\includegraphics[width=0.125\textwidth]{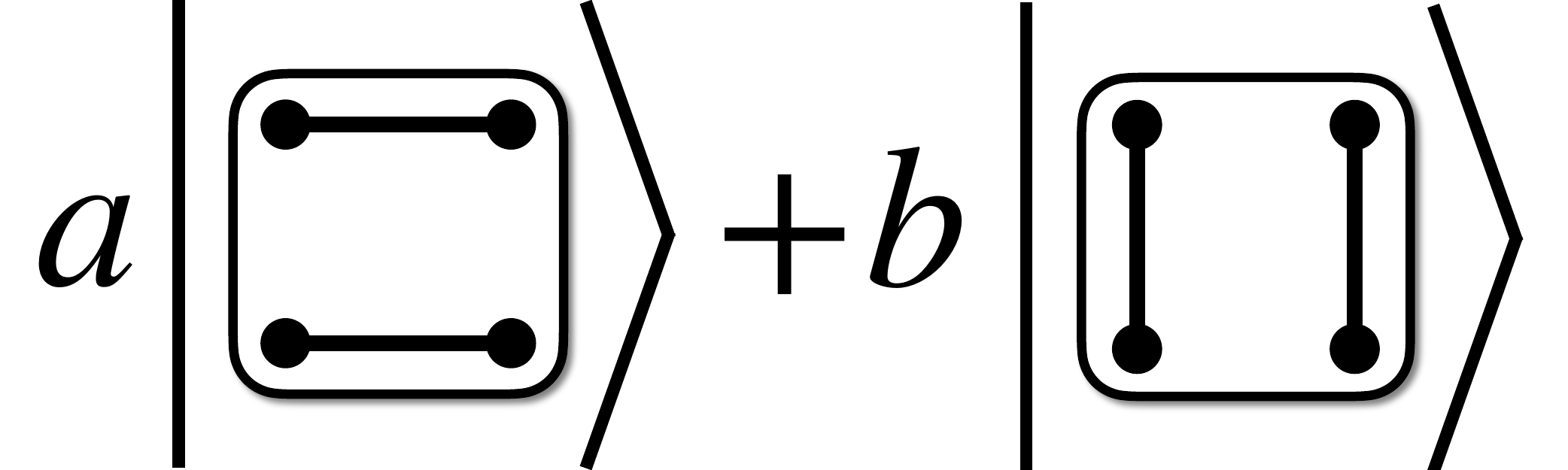}
}}
\includegraphics[width=0.32\textwidth]{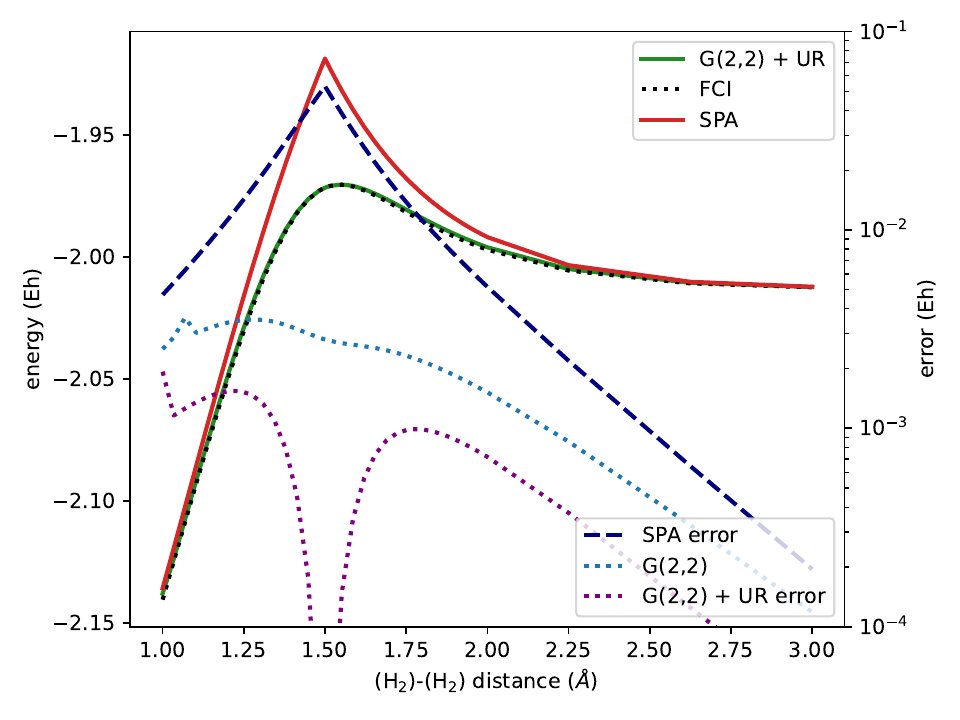}
\shiftleft{3.25cm}{\raisebox{1.3cm}[0cm][0cm]{
\includegraphics[width=0.125\textwidth]{h4-resonance.pdf}
}}
    \caption{{Total energy error of the SPA method (with optimized orbitals) for H$_6$/STO-6G(6,12) and the rectangular H$_4$/STO-6G(4,8) with the used Lewis structure depicted in the plots. Left: linear H$_6$ with simultaneous  stretching of all H-H distances. Middle: square H$_4$ with simultaneous  stretching of all H-H distances. Right: rectangular H$_4$ with two H$_2$ units (fixed H-H distance of 1.5\AA) and varying inter-molecular distance.  The used SPA circuits are identical to the ones in Tab.~1 and~2 of Ref.~\citenum{kottmann2022molecular} having total depth of 3 and 3, respectively 2, independent parameters. Note that the linear H$_4$ (not depicted) behaves similar as the linear H$_6$. In addition, the G(2,2)+UR energies are shown for the rectangular H$_4$ in two different scenarios.}}\label{fig:h6-spa}
\end{figure*}

A further illustration of the weak type of correlation contributed by the third graph, is to take a wavefunction generated by the first two graphs, but with more flexibility in the individual circuits. In this case we added more orbital rotations to the circuit (denoted by $U_\text{R}$ in the corresponding methods), so that all non-connected vertices of the graphs were connected through an orbital rotation. {With this additional expressibility, the $G(2,2)+U_\text{R}$ wavefunction is sufficient to represent the true ground state accurately (see Fig.~\ref{fig:h6-spa}), therefore following Rumer's selection rules ~\cite{wu2011classicalvbt} for valence bond structures that exclude ``crossed'' structures such as the third graph in Fig.~\ref{fig:graph_basis_construction}. Note that the ``cusp''~\cite{paldus1993appliation} at the transition between rectangular and quadratic H$_4$ vanishes already for G(2,2).}

\begin{figure*}
    \centering
    \raisebox{-0.225cm}[0cm][0cm]{\includegraphics[height=0.24\textheight]{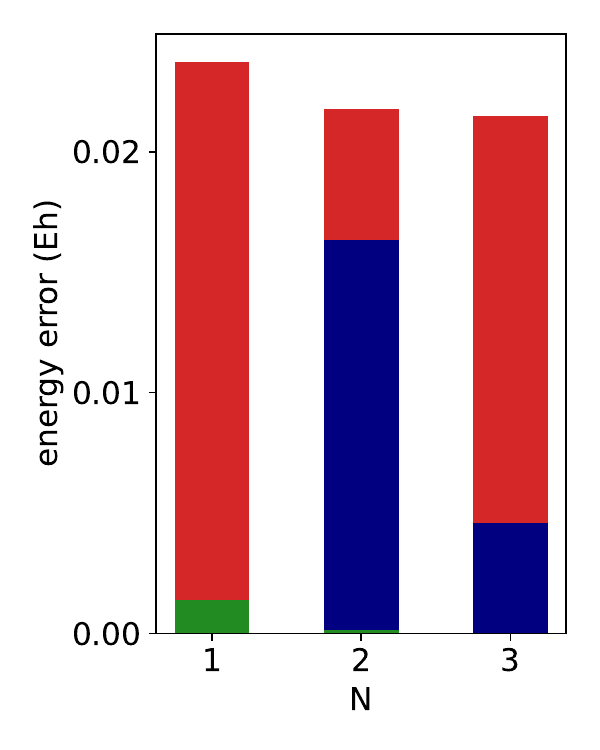}}
    \shiftleft{4.5cm}{\raisebox{5.5cm}[0cm][0cm]{(a)}}
    \shiftleft{3.2cm}{\raisebox{5.8cm}[0cm][0cm]{H$_4$, $r=1.5$\AA}}
    \hspace{0.25cm}
   {\includegraphics[height=0.25\textheight]{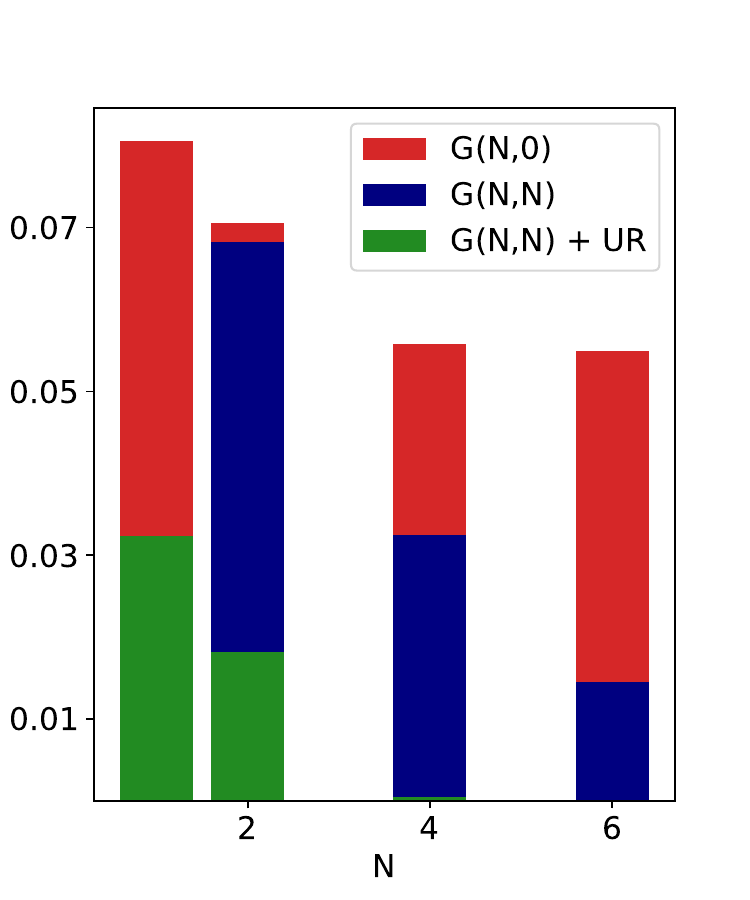}}
    \shiftleft{5.5cm}{\raisebox{5.6cm}[0cm][0cm]{(b)}}
    \shiftleft{3.8cm}{\raisebox{5.8cm}[0cm][0cm]{H$_6$, $r=1.5$\AA}}
    \hspace{0.25cm}
    \raisebox{0.5cm}[0cm][0cm]{\includegraphics[width=0.2\textwidth]{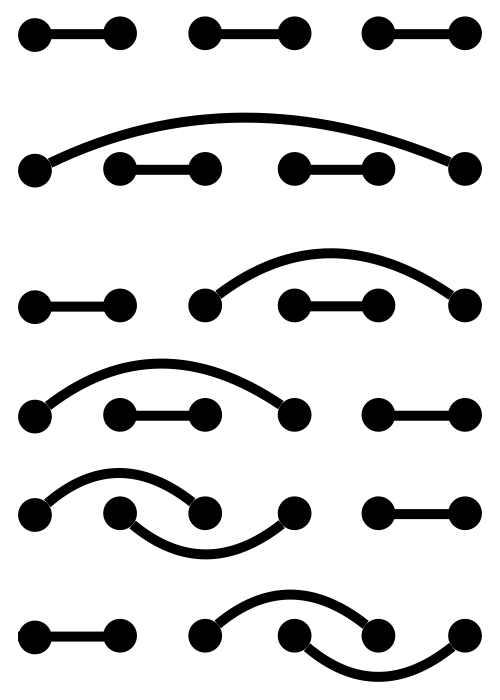}}
    \shiftleft{4.5cm}{\raisebox{5.6cm}[0cm][0cm]{(c)}}
    \caption{Effects of optimization level and circuit expressivity on linear hydrogen chains (equidistant bond lengths of 1.5\AA). Basic circuits are constructed analogously to Fig.~\ref{fig:graph_basis_construction}. For $G(N,) + UR$ datapoints the circuits were augmented with additional orbital rotations. (a): H$_4$/STO-6G(4,8)  (b): H$_6$/STO-6G(6,12), (c): graphs used for H$_6$/STO-6G(6,12). {We recall that $G(1,0)$ and $G(1,1)$ are equivalent, this is why $G(1,1)$ is absent in panel (b), while $N=3,5$ are missing because we employ degenerate graphs together (there are two degenerate couples). Hence graphs 3 and 4, respectively 5 and 6, were added in groups to account for their structural symmetries.}}
    \label{fig:h6-effects-of-ur}
\end{figure*}

\begin{table}[]
    \centering
    \begin{tabular}{lcc}
        \toprule
        Method &  H$_4$ & H$_6$ \\
        \midrule
        MRSQK ($m=1$) &  2656 & 19944 \\
        MRSQK ($m=8$) &  21248 & 159552 \\
        \midrule
        UpCCGSD & 188 & 626\\
        2-UpCCGSD & 432 & 1387 \\
        \midrule
        SPA & 6 & 9 \\
        SPA+ & 116 & 197 \\
        SPA+X & 294 & 489 \\
        \midrule
        $G(N,M)$ & 70 & 150  \\
        $G(N,M)+U_\text{R}$ & 150 & 425\\
        \bottomrule
    \end{tabular}
    \caption{\textsc{CNOT} counts of the deepest circuit in MRSQK {with 1 and 8 Trotter steps ($m=1,8$)} , $k$-UpCCGSD (compiled with optimizations introduced in Ref.~\citenum{kottmann2022optimized}), as for example used in NO-VQE, {SPA+(X) where a single circuit is constructed from the two leading graphs, and the $G(N,M)$ developed in this work. See Ref.~\citenum{kottmann2022molecular} for $k$-UpCCGSD and SPA+(X) performance on the two systems.} Note further reduction could be achieved through more efficient compiling of the $U_\text{R}$ rotations ~\cite{kivlichan2018quantum, google2020hartree}}
   \label{tab:cnot_counts}
\end{table}

\begin{figure*}
    \centering
    \includegraphics[width=0.45\textwidth]{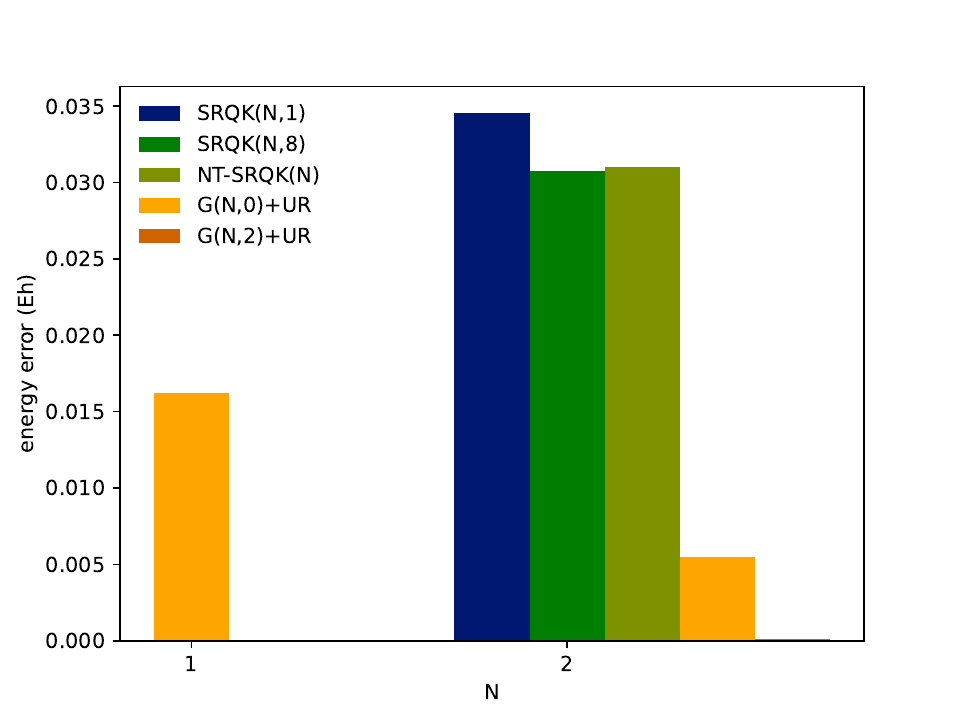}
    \shiftleft{8.25cm}{\raisebox{4.6cm}[0cm][0cm]{(a)}}
    \includegraphics[width=0.45\textwidth]{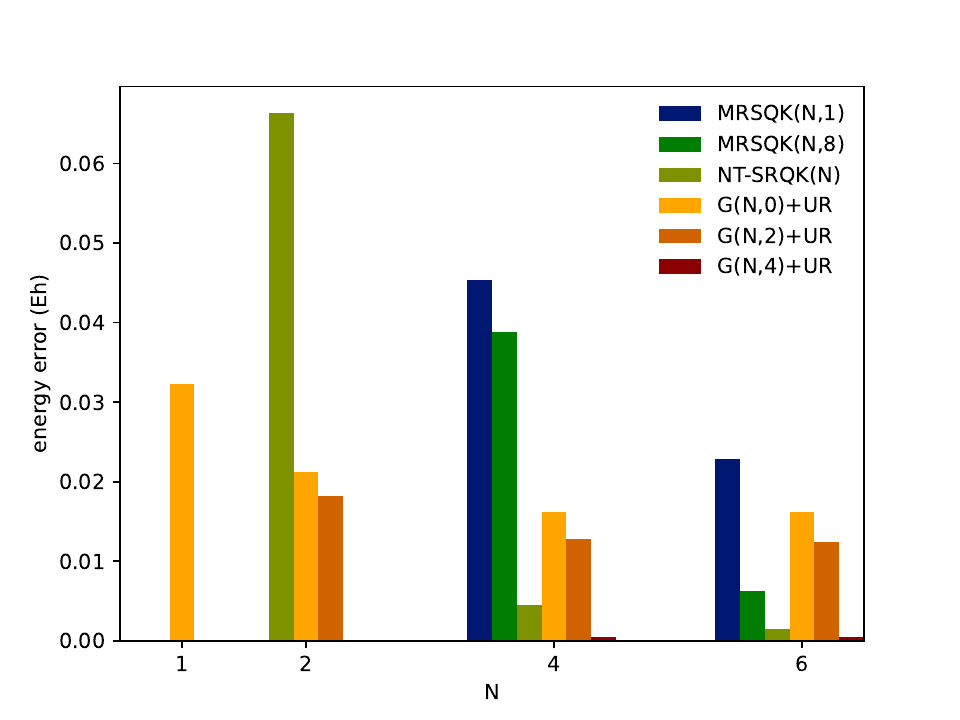}
    \shiftleft{8.25cm}{\raisebox{4.6cm}[0cm][0cm]{(b)}}
    \caption{Comparison of the present approach with the multireference selected quantum Krylov (MRSQK), as well as the single reference quantum Krylov (SRQK) method~\cite{stair2020multireference} with respect to the number $N$ of used effective many-body basis states. Energy errors with respect to the exact ground state are given for (a) H$_4$/STO-6G(4,8) and (b) H$_6$/STO-6G(6,12) -- same molecules as in Fig.~\ref{fig:h6-effects-of-ur}. MRSQK($N,m$) and NT-SRQK($N$) results are with, and without, Trotter approximation using $m$ Trotter steps for the real-time evolution. For MRSQK the number of reference states was fixed to 2, so that non-trivial results arise at $N>2$ (similar for SRQK with $N>1$). Note that NT-SRQK is an independent implementation. Graph based construction is done according to Fig.~\ref{fig:graph_basis_construction} with static energies {($G(N,0)$)} denoting the ground state of the effective Hamiltonian in the pre-optimized basis. $G(N,M)$ denotes energies of wavefunctions according to Eq.~\eqref{eq:concerted-opt} using $N$ graphs to construct $N$ circuits $U_k$ with $M$ of them being fully optimized in a concerted optimization.}
    \label{fig:mrqk_comparison}
\end{figure*}

\subsection{Linear H$_4$ and H$_6$: comparison to quantum Krylov}

In the previous Section, we have seen how the individual parts of the wavefunction can be interpreted. We have identified weak correlations (as in the third graph of the rectangular H$_4$) that can not be generated through energy based pre-optimization as the energetic effect is due to destructive interference and only present in the total wavefunction. In the case of the rectangular H$_4$ those weak correlations could be compensated by equipping the circuits representing the individual graphs with more freedom in the form of orbital rotations. We expect a similar behavior for other systems and tested it on the linear H$_4$ and H$_6$ models where the results displayed in Fig.~\ref{fig:h6-effects-of-ur} show the same trends as observed before. While the non augmented wavefunctions show good convergence within the first 3, respectively 6, graphs, the overall error is still around 5, respectively 10, millihartree for the fully optimized wavefunctions. On the other hand, the augmented $G(N,M=N)+U_\text{R}$ wavefunction already achieves chemical accuracy at a smaller size $N$ of the effective many-body basis.\\

The linear H$_4$ and H$_6$ hydrogen chains are prominent benchmark systems that have been applied in the context of  Ref.~\citenum{kottmann2022molecular} as well as in Ref.~\citenum{stair2020multireference} that introduced the Multi-Reference Selected Quantum Krylov (MRSQK) method -- a real-time evolution approach towards approximating the Krylov subspace in Eq.~\eqref{eq:krylov}. 
{In particular, the MRSQK method starts with $d$ selected Slater determinants instead of a single initial state $\ket{\psi_0}$ to generate the non-orthogonal Krylov subspace. Additional effective basis states are constructed by applying the time evolution operator generated by the molecular Hamiltonian in Eq.~\eqref{eq:hamiltonian} leading to $N = d(s + 1)$ Krylov basis states. In order to implement this on a quantum processor, Trotterization with different number of Trotter steps ($m$) is necessary. 
In Fig.~\ref{fig:mrqk_comparison} we compare the $G(N,M)+U_\text{R}$ energies with MRSQK($N,m$) and SRQK($N,m$) with respect to the size of the many-body basis $N$ that is, the number of states in the Krylov basis. SRQK corresponds to MRSQK with a single initial state (the Hartree-Fock determinant) and NT-SRQK to a non Trotterized simulation ($m \rightarrow \infty$) -- note that $m$ and $M$ are not related. For H$_6$ we set $d=2$ and used $s\in\{0,1,2\}$ for the $N=2,4,6$ energies. For the (NT-)SRQK results $N=s$ and $d=1$ by definition.} We see that {$G(N,N)+U_\text{R}$} always outperforms MRSQK in all flavors while $G(N>2,2)$ energies can not improve upon the non-Trotterized quantum Krylov variant and $G(N>4,0)$ can not improve upon the Trotterized variant with 8 Trotter steps. Based on the observations form the previous section, this is not further surprising as we would expect the higher order graphs only to bring significant improvements when they are included into the concerted optimization. Note that apart from the basis size $N$ the $G(N,M)$ method requires significantly shallower circuits compared to the Trotterized real-time evolutions necessary to generate the MRSQK basis (see Tab.~\ref{tab:cnot_counts}). In comparison to NO-VQE~\cite{huggins2020non} the circuit sizes are still significantly reduced and the method in this work is not relying on repeated randomized initialization. The total number of BFGS iterations is moderate (varying between 15 and 30 iterations) and we expect further reduction through improved implementations.
 
\begin{figure*}
\includegraphics[width=0.45\textwidth]{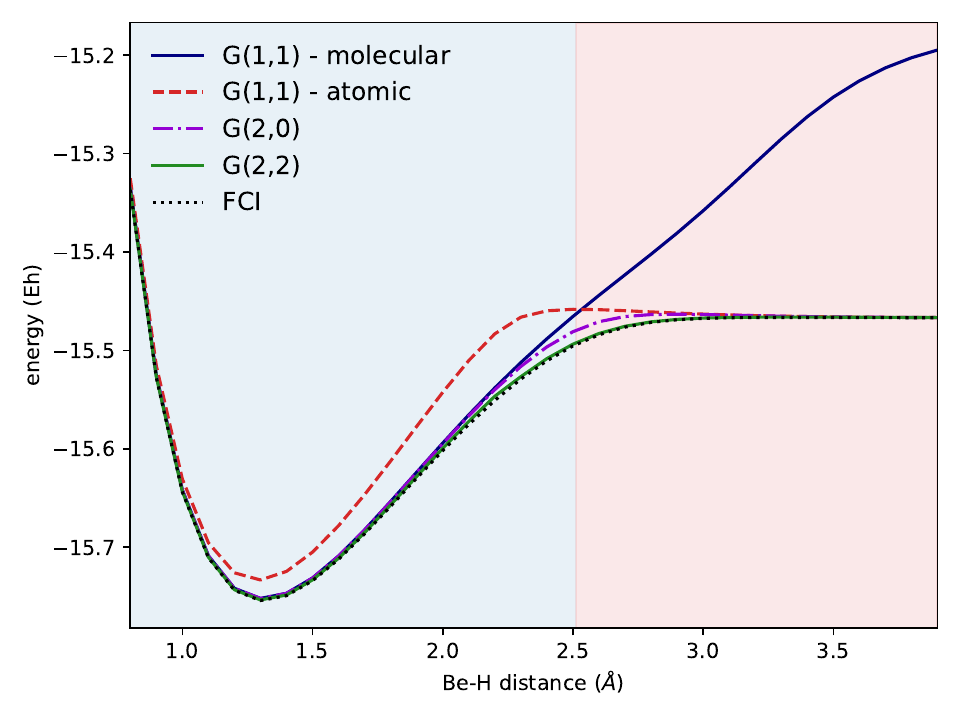}
\shiftleft{8.5cm}{\raisebox{4.6cm}[0cm][0cm]{(a)}}
\shiftleft{6.8cm}{\raisebox{2.3cm}[0cm][0cm]{
\includegraphics[width=0.1\textwidth]{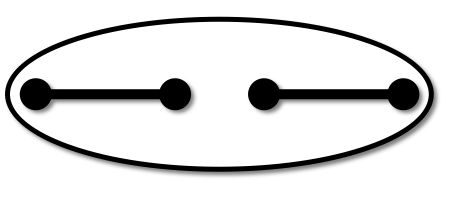}
}}
\shiftleft{3.1cm}{\raisebox{2.3cm}[0cm][0cm]{
\includegraphics[width=0.1\textwidth]{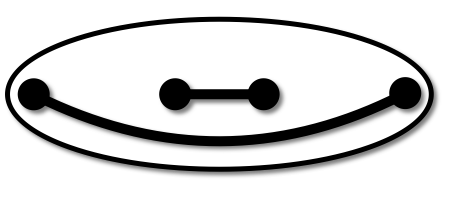}
}}
\includegraphics[width=0.45\textwidth]{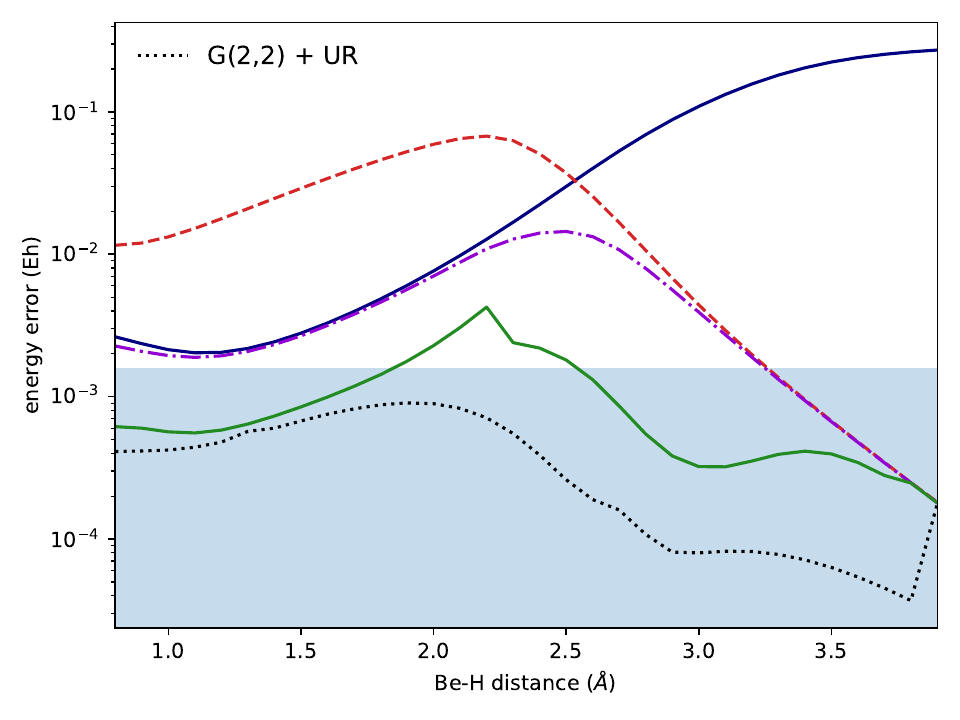}
\shiftleft{8.5cm}{\raisebox{4.6cm}[0cm][0cm]{(b)}}
\caption{BeH$_2$/STO-6G(4,8) energies from $G(N,M\leq N)$ wavefunctions constructed analog to Fig.~\ref{fig:graph_basis_construction}. (a): absolute energies with dominant molecular graphs shown as insets and corresponding regions highlighted in color, (b): energies relative to {the Full Configuration Interaction (FCI)}/STO-6G(4,8) with the 1~kcal/mol accuracy region highlighted in blue. $G(2,2) + U_R$ denotes the same procedure with additional orbital rotators in the individual circuits.}\label{fig:beh2_8q}
\end{figure*}

\subsection{Linear BeH$_2$: transferring concepts from H$_4$ to a more complex system}
In Ref.~\citenum{kottmann2022molecular} the graph based construction was applied for a single circuit to prepare the wavefunction directly, while in this work we resorted to a divided approach where each graph corresponds to an individual circuit. In the previous section we have seen, that this approach can achieve comparable accuracy in the wavefunctions.
So far, we resorted to simplified hydrogenic systems with a single spherical s-type orbital on each atom.
With BeH$_2$/STO-6G(4,8) we add a model system with more complicated orbital structure (having s- and p-type orbitals on the central Be atom). This model system is the same as in Ref.~\citenum{kottmann2022molecular}, {with p$_x$ and p$_y$ removed and the lowest orbital from the Hartree-Fock method frozen with double occupancy (frozen-core),} and has the same dimensions as H$_4$/STO-6G(4,8). Through the graph description we can treat the BeH$_2$ now in the same way as the linear H$_4$. The two main graphs are illustrated in Fig.~\ref{fig:beh2_8q}, where one of them is interpreted as molecular and the other as atomic (see also Eq.~(25) in Ref.~\citenum{kottmann2022molecular}). In Fig.~\ref{fig:beh2_8q} we clearly see how the potential energy surface is divided into three domains: the first being the bonded domain (around bond distance $R=1.5$\AA), the second the dissociated domain ($R>3.0$\AA) both dominated by a single graph, while the third domain (around $R=2.6$\AA) requires both graphs for an accurate description.

\section{Conclusion \& Outlook}
{In this work, we developed an effective basis method for multiconfigurational valence bond wavefunctions on quantum computers.}
We could show, how this method outperforms other effective basis methods in terms of size of the effective basis as well as individual quantum resources for the involved circuits. 
Most importantly the developed method allows us to interpret the results and to learn from the discovered effects.\\
At this point, the computational bottleneck comes with the concerted optimization necessary for the determination of the effective basis. As a quantum algorithm, this procedure requires many evaluations of primitive expectation values (in the sense of Ref.~\citenum{tequila}). We see however promising ways forward in that respect, as illustrated in the following.
Through the combination with the circuit designs of Ref.~\citenum{kottmann2022molecular} the effective basis is described by individual circuits that are equivalent to separable pair approximations (SPA)~\cite{kottmann2022optimized}. As a consequence, the individual wavefunctions are classically simulable, so that energy based pre-optimization can be performed purely classical with linear memory requirement. The $G(N,0)$ method therefore defines a de-quantized and de-randomized, both with respect to parameter determination, flavor of a NO-VQE. Based on the reported numerical evidence, we expect this to work well for qualitative descriptions of the wavefunctions. For a quantitative treatment, energy based pre-optimization is however not expected to be practicable. Other types of pre-optimizations, as in Ref.~\citenum{baek2022sayno}, could however be imagined for the future and might lead to more powerful purely classical methods capable of generating compact quantum circuits for accurate state preparation. 
The interpretable circuit design offers here a chance to effectively predict optimal circuit parameters based on detailed analysis of model systems. {In general, getting all possible Lewis structures as employed in this work would correspond to getting all perfect matchings of a fully connected graph -- an untractable task. The hope is, that the necessary number of graphs for a decent approximation of the target wavefunction remains tractable for a wide range of usecases. The linear H$_6$ molecule provides an illustrative example, where we demonstrated sufficient accuracy with 4 graphs. In this work, the graphs were manually selected, the obtained results and established connections to valence bond theory are however a promising prospect for the development of automated methods.}

\section{Acknowledgment}
This project was initiated through the Mentorship Program of the Quantum Open Source Foundation (QOSF) Cohort~5. We are grateful to the QOSF for providing this platform. 
JSK likes to thank Stijn De Baerdemacker, Ehsan Ghasempouri, and Kjell Jorner for insightful discussions on the connections to valence bond theory as well as Oriol Vendrell and Garnet Chan for pointing out connections to GVB in prior works. Special thanks to Kjell for confirming that VBSCF is indeed the closest relative to G(N,M) and providing some additional references.
We thank Philipp Schleich for giving valuable feedback to the initial manuscript. Furthermore we thank Nick Stair and Francesco Evangelista for providing user-friendly public access to the MRSQK~\cite{stair2020multireference} method through \textsc{qforte}~\cite{stair2022qforte} as well as all open-source developers of the used packages and dependencies.

\bibliography{main.bib}
\clearpage
\appendix
\onecolumngrid

\section{Explicit qubit states for the graphical depictions in the main text}
In the main text we resorted to graphical symbols to represent electronic configurations in the rectangular H$_4$ system. Here we show the corresponding configurations as qubit states in Jordan-Wigner encoding - meaning they are identical to standard occupation number vectors in second quantized formulation.
The qubit represent spin orbitals (even qubits spin-up and odd qubits spin-down) and the orbitals are orthonormalized atomic basis orbitals of the STO-6G set - meaning one spherical symmetrical s-type orbital for each of the four H atoms in clockwise order. The four configurations featuring four electrons are

\begin{align}
\ket{\raisebox{-0.3cm}{\includegraphics[width=0.045\textwidth]{configuration.008.png}}} =& \ket{11000011}, &
\ket{\raisebox{-0.3cm}{\includegraphics[width=0.045\textwidth]{configuration.009.png}}} =& \ket{11110000}, \\ 
\ket{\raisebox{-0.3cm}{\includegraphics[width=0.045\textwidth]{configuration.010.png}}} =& \ket{00111100}, & 
\ket{\raisebox{-0.3cm}{\includegraphics[width=0.045\textwidth]{configuration.011.png}}} =& \ket{00001111}.
\end{align}

The two electron states represent the configuration on a reduced orbital set given by the edge of the graph
\begin{align}
\ket{\raisebox{-0.3cm}{\includegraphics[width=0.045\textwidth]{configuration.004.png}}} =& \ket{1100}_{0145}, &
\ket{\raisebox{-0.3cm}{\includegraphics[width=0.045\textwidth]{configuration.005.png}}} =& \ket{0011}_{2367}, \\
\ket{\raisebox{-0.3cm}{\includegraphics[width=0.045\textwidth]{configuration.006.png}}} =& \ket{0011}_{0145}, &
\ket{\raisebox{-0.3cm}{\includegraphics[width=0.045\textwidth]{configuration.007.png}}} =& \ket{1100}_{2367},
\end{align}
Here, the subscripts denote the qubit indices in order for the result of the tensor products in the main text to be in the right order. This is meant in the following way:
\begin{align}
\ket{\raisebox{-0.3cm}{\includegraphics[width=0.045\textwidth]{configuration.004.png}}} \otimes \ket{\raisebox{-0.3cm}{\includegraphics[width=0.045\textwidth]{configuration.005.png}}} =& \ket{1100}_{0145} \otimes \ket{0011}_{2367} = \ket{11000011} = \ket{\raisebox{-0.3cm}{\includegraphics[width=0.045\textwidth]{configuration.008.png}}}
\end{align}

\section{Explicit code example}\label{sec:code-example}
Further down, explicit code examples to reproduce the data in Fig.~\ref{fig:graph_basis_construction} are provided. The code can be viewed as pseudocode, it is however executable with \textsc{tequila}~\cite{tequila} (version 1.8.4) and \textsc{numpy}~\cite{numpy}(version 1.21.5). The code requires \textsc{psi4}~\cite{psi4} or \textsc{pyscf}~\cite{pyscf1,pyscf2} to be installed (in order to compute the molecular integrals) and it is recommended to have \textsc{qulacs}~\cite{qulacs} installed as wavefunction simulation backend. In the first block, the $G(N,M)$ function is defined. Note that we used a different implementation that exploits some shortcuts in the classical simulation - this is provided on an external Github repository~\cite{github-repo}.\\

 The function $G(N,M)$ can be implemented as:
\clearpage
\begin{lstlisting}[language=Python]
# tq.version.__version__ >= 1.8.5
import tequila as tq
import numpy

def G(N,M,H,U,initial_values):
    print(N," ",M)
    assert M <= N
    assert N <= len(U)

    # initial values for coefficients through GEM (Eq.1)
    v,vv = tq.apps.gem(U, H, variables=initial_values)
    
    # initialize variables for coefficients
    c = [tq.Variable(("c",i)) for i in range(N)]
    
    for i in range(N):
        initial_values[c[i]]=vv[i,0]

    # Initialize the objective of Eq.7 
    # by expanding with wavefunction of Eq.6
    ED = 0.0
    EN = 0.0
    for i in range(N):
        for j in range(N):
            # only need real part of braket (by construction)
            EN += c[i]*c[j]*tq.braket(U[i],U[j],H)[0]
            ED += c[i]*c[j]*tq.braket(U[i],U[j])[0]
    
    E = EN/ED
    
    # active variables for the concerted optimization
    active = sum([U[i].extract_variables() for i in range(M)],[])
    active += c

    # gradient compilation not optimized for tq.braket
    # recommending to use finite-differences
    result = tq.minimize(E, variables=active, initial_values=initial_values, gradient="2-point", method_options={"finite_diff_rel_step":1.e-4})
    
    return result
\end{lstlisting}
 with this function we can compute the data shown in Fig.~\ref{fig:graph_basis_construction} in the main text as
\begin{lstlisting}[language=Python]
# define the molecule
geometry = "H 1.5 0.0 0.0\nH 0.0 0.0 0.0\nH 1.5 0.0 1.5\nH 0.0 0.0 1.5"
mol = tq.Molecule(geometry=geometry, basis_set="sto-6g")
# use orthonormalized atomic orbitals
mol = mol.use_native_orbitals()

# Define the qubit Hamiltonian
H = mol.make_hamiltonian()
# get the reference energy
exact = mol.compute_energy("fci")

# define the graphs
G1 = [(0,1),(2,3)]
G2 = [(0,2),(1,3)]
G3 = [(0,3),(1,2)]
graphs = [G1, G2, G3]

# define the circuits to generate the basis
circuits = []
for i,edges in enumerate(graphs):
    # label to prevent identical variable names in different circuits
    label="G{}".format(i+1)
    U = mol.make_ansatz(name="SPA", edges=edges, label=label)
    for e in edges:
        U += mol.UR(e[0],e[1],label=label) 
    circuits.append(U)

# pre-optimize the circuits
variables = {}
for U in circuits:
    E = tq.ExpectationValue(U=U, H=H)
    result = tq.minimize(E, silent=True)
    variables = {**variables, **result.variables}
    print("Graph {}: with energy {:+2.5f}".format(i, result.energy))

# run the G(N,M) computations
energies={}
for N in range(1,4):
    for M in range(1,N):
        result = G(N,M,H,circuits,variables)
        variables = result.variables
        energies["G({},{})".format(N,M)]=result.energy
\end{lstlisting}
\clearpage
As it is often useful to employ SPA circuits in order to identify useful test and benchmark systems, we also provide the full script to reproduce the H$_6$ data from Fig.~\ref{fig:h6-spa}:
\begin{lstlisting}[language=Python]
# tq version 1.8.9
# qulacs version 0.6.1
# pyscf version 2.1.1
import tequila as tq
from tequila.quantumchemistry import optimize_orbitals
import numpy

geometry = """
H 0.0 0.0 {}
H 0.0 0.0 {}
H 0.0 0.0 {}
H 0.0 0.0 {}
H 0.0 0.0 {}
H 0.0 0.0 {}
"""

guess=numpy.eye(6)
guess[0]=[1.0,1.0,.0,.0,.0,.0]
guess[1]=[1.0,-1.0,.0,.0,.0,.0]
guess[2]=[.0,.0,1.0,1.0,.0,.0]
guess[3]=[.0,.0,1.0,-1.0,.0,.0]
guess[4]=[.0,.0,.0,.0,1.0,1.0]
guess[5]=[.0,.0,.0,.0,1.0,-1.0]
guess = guess.T

error = {}
exact = {}
spa = {}
for R in numpy.linspace(0.5,4.5,50):

    geom = geometry.format(0.0,R,2*R,3*R,4*R,5*R)
    mol = tq.Molecule(geometry=geom, basis_set="sto-3g").use_native_orbitals()
    fci = mol.compute_energy("fci")
    U = mol.make_ansatz(name="HCB-SPA", edges=[(0,1),(2,3),(4,5)])

    opt = optimize_orbitals(mol, U, use_hcb=True, initial_guess=guess)
    guess = opt.mo_coeff
    print(fci-opt.energy)
    error[R]=fci-opt.energy
    exact[R]=fci
    spa[R]=opt.energy

print("fci=",exact)
print("spa=",spa)
print("error=",error)
\end{lstlisting}
\clearpage

\section{Details on related works}\label{sec:related-works}
In the introduction of the main text, several methods that try to approximate the Krylov subspace in Eq.~\eqref{eq:krylov} where grouped together. These works take different routes to approximate the $\ket{\psi_k} \in \mathcal{K}$. In Refs.~\citenum{yeter2020practical,motta2020determining, tsuchimochi2022multistate} Quantum Imaginary Time evolution (QITE) is employed while Ref.~\citenum{stair2020multireference} resorts to real-time evolution on multiple initial states defining a Multi-Reference selected Quantum Krylov (MRSQK) approach. Similar to MRSQK are the quantum filter diagonalization (QFD)~\cite{parrish2019quantum2, cohn2022quantum} and the variational assisted quantum simulator (QAS)~\cite{bharti2021quantum, bharti2021iterative}. The QFD relies on real-time simulation in the same spirit as MRSQK while QAS approximates powers of the Hamiltonian by creating products of  individual unitaries (Pauli strings) that define it.  In a similar fashion, an inverse power method (using $H^{-k}$) has been proposed with analogue quantum simulators in mind.~\cite{kyriienko2020quantum} Recently, direct unitary encoding of the Hamiltonian powers was proposed in Ref.~\citenum{seki2021quantum} (via linear combination of unitaries) and~\cite{kirby2023exact} (via block-encoding). \\
Related to the non-orthogonal variational quantum eigensolvers (NOVQE~\cite{huggins2020non}) are NOVQEs with classical preoptimized parameters (NOQE~\cite{baek2022sayno}), variational quantum subspace expansion \cite{mcclean2020decoding,takeshita2020increasing, yoshioka2022generalized,urbanek2020chemistry} and entirely classical approaches like NOMAGIC~\cite{mcclean2015nomagic}. The NOQE method in particular introduces an interesting alternative for pre-screening of parameters.\\

\end{document}